\def\vday{March 25, 1994}
\documentstyle[preprint,aps]{revtex}
\begin{document}
\def\check{*\marginpar{$\leftarrow$\tiny{\vday}} }
\draft
\pagestyle{empty}
\centerline{
\hfill NHCU--HEP--94/11-T}\vskip -.3cm
\centerline{\Large Transfer Matrix Method in Sandpile Models}\vskip 1cm
\centerline{
Darwin Chang$^{(1)}$
Chen-Shan Chin$^{(1)}$ and
Shih-Chang Lee$^{(2)}$ }\vskip 1cm
\centerline{
$^{(1)}$Physics Department,
National Tsing-Hua University, Hsinchu, Taiwan}
\centerline{
$^{(2)}$Institute of Physics, Academia Sinica, Taipei, Taiwan}
\date{\today}

\begin{abstract}
We present a transfer matrix method which is particularly useful
for solving some classes of sandpile models.  The method is then used to
solve the deterministic nonabelian sandpile models for N=2 and N=3.
The possibility of generalization to arbitrary N is discussed briefly.
\end{abstract}
\vfill
%
%
\narrowtext
\pagestyle{plain}
%
%
\section{Introduction}
Sandpile model is the simplest model of self-organized
criticality(SOC). This kind of models were first discussed by Bak,
Tang, and Wiensenfeld\cite{BTW}. They used numerical method to study
some of the sandpile models and observed that the models automatically
evolve into a self-organized critical state when they possess $1/f$
spectra in both spatial and temporal distributions of certain physical
quantities.  They suspected that SOC maybe an universal characteristic
underlying the the nonlinear dispersive systems such as earth quakes,
forest fire, turbulence, etc, which are prevalent in nature.

The simplest sandpile model is a cellular automata on an one
dimensional lattice with a height number $h_i$ assigned to each site.
There are two basic operations of the model \-- dropping and toppling.
Dropping means that one sand is added at some site of the lattice,
i.e. $D_i: h_i\rightarrow h_{i+1}$.  Toppling occurs when a slope
(defined as the difference in height between adjacent sites) exceeds
some critical value. If toppling occurs at one site, some sands at the
site will be moved to the other sites which may trigger further
topplings.

Though the rules of evolution of sandpile model are typically very
simple, it is hard to solve them analytically when the degrees of
freedom become very large. Most researcher handle the models by
numerical simulation\cite{NS1}\cite{NS2}\cite{NS3}.  If the rules are
such that the evolution of the system is independent of the order of
the droppings, then the model is called abelian.  For a large class of
abelian models some exact results have been obtained by Dhar et
al.\cite{abelian}  The non-abelian ones are harder to solve and there
exists little exact result. In a previous work \cite{Lee1}, one of
the authors investigated a class of non-abelian sandpile models and
was able to solve the model in the deterministic case when the sand is
dropped at a fixed site.  In this note, we wish to present a new method
of solving this class of sandpile models. We consider the one
dimensional case and label the sites from left to right as 1 to L.
The sand is dropped only at the site 1. If the slope at a site exceeds
a given number $N$, then the sand will topple to the right. Let the slope
$\sigma_i=h_i-h_{i+1}$, then the toppling rule is ``if $\sigma_i>N$,
then $\sigma_{i-1}\rightarrow\sigma_{i-1}+N$,
$\sigma_i\rightarrow\sigma_i-(N+1)$ , and
$\sigma_{i+N}\rightarrow\sigma_{i+N}+1$''. The rule should be modified
when toppling occurs near the boundary.  The condition at the left
boundary is trivial.  When sands reach beyond the right boundary they drop
out from the system, i.e. we keep $h_i=0$ for $i>L$. A state in which
all $\sigma_i\leq N$ is called a stable state.  Toppling stops when a
stable state is reached.  Each dropping and subsequent toppling
processes will result in the transition from one stable state to
another.  Since there are only a
finite number of different states in the system, after dropping enough
sand at site 1, the system will step into a cycle called the limit
cycle.  For the system that we consider here there is only one limit
cycle in the problem.  The number of different states in the limit
cycle is $N^L$.  In the following, all the states we will refer to
are the stable states.

The method to be introduced here is particularly
useful for those models in which the
structure of the limit cycles has been work out.  Once the cycle structure
is known, its information can be succinctly summarized in a matrix
which we call the transfer matrix by analogy with the similar matrix in
statistical mechanics.  For the deterministic model defined above, the
structure of the limit cycle has been worked out in ref.\cite{Lee1}.
Therefore we should use it as our main example, the method may be
applicable to a much wider classes of models.

\section{The definition of Transfer matrix}
A state is in the limit circle if and only if
\begin{enumerate}
\item There exists at least one site $i$ for any consecutive $N$ sites
  such that $\sigma_i=N$.\label{enum:constrain1}
\item There exists a site $i$ satisfying $L-\sigma_L\leq i < L$ such
  that $\sigma_i=N$.\label{enum:constrain2}
\end{enumerate}
Since the first condition is for $N$ consecutive sites, there is no
constraint in consecutive $N-1$ sites. Choose the complete set of
states for consecutive $N-1$ sites as a basis. For example in the $N=2$
model, the states in the basis is $|0\rangle,\;|1\rangle,\;|2\rangle$.
For $N=3$ model, the basis is of the form $|33\rangle,
\;|a3\rangle,\;|3a\rangle,\;|ab\rangle $, where
$a,\;b$ belongs to $\{0,1,2\}$. For arbitrary $N$, there are
$\left( N+1 \right)^{(N-1)}$ states in the basis. Define the transfer
matrix as a mapping between
$|\eta_i\rangle\equiv|\sigma_i\sigma_{i+1}\cdots\sigma_{i+N-2}\rangle$
and
$|\eta_{i+1}\rangle\equiv|\sigma_{i+1}\sigma_{i+2}\cdots\sigma_{i+N-1}\rangle$
such that
\begin{equation}
\langle\eta_i|T|\eta_{i+1}\rangle=\left\{
  \begin{array}{l}
    1\;\;\mbox{if the sequence $\sigma_i\sigma_i+1\cdots\sigma_{i+N-1}$
satisfies constrain \ref{enum:constrain1}} \\
    0\;\;\mbox{otherwise}
  \end{array}.
  \right.
\end{equation}

For example, in the $N=2$ model, if $\sigma_{i+1}=2$, the $\sigma_i$
could be $0,\;1,\;2$, so
$\langle0|T|2\rangle=\langle1|T|2\rangle=\langle2|T|2\rangle=1$.  If
$\sigma_{i}=2$, then $\sigma_{i+1}$ could be $0,\;1,\;,2$, so that
$\langle2|T|0\rangle=\langle2|T|1\rangle=1$. For other cases, $\langle
f|T|i\rangle=0$.

Let $|0\rangle=\mbox{col}(1,0,0)$, $|1\rangle=\mbox{col}(0,1,0)$,
$|2\rangle=\mbox{col}(0,0,1)$, then we can write the transfer matrix
in the following form:
  \begin{equation}
    T=\left[
     \begin{array}{ccc}
     0&0&1 \\
     0&0&1 \\
     1&1&1
    \end{array}
   \right].
  \end{equation}
Then the number of states starting with
$|\sigma_i\cdots\sigma_{i+N-2}\rangle$, and ending with
$|\sigma_j\cdots\sigma_{j+N-2}\rangle$ in one limit circle is
$\langle\sigma_i\cdots\sigma_{i+N-2}| T^{(j-i)}
|\sigma_j\cdots\sigma_{j+N-2}\rangle$.

The transfer matrix contains all the information about the time
average of space dependent functions, such as the one point function
$\langle\sigma_i\rangle$, and the two point function
$\langle\sigma_i\sigma_j\rangle$. Here $\langle\cdots\rangle$ means the
average in time over a limit circle.
Define
${\cal I} =\{$ the complete set of states
$|\sigma_{L-N+2}\cdots\sigma_{L}\rangle$ allowed by the boundary condition
  (\ref{enum:constrain2}) $\}$,
${\cal A} =\{$ complete set of states
$|\sigma_1\cdots\sigma_{N-1}\rangle$ in general $\}$.
The one point function can be evaluated in terms of the summation of all states
in a limit circle, i.e.
\begin{equation}
\langle\sigma_i\rangle = \frac{1}{N^L} \sum_{\eta_i}
  \sum_{\alpha \in {\cal A}} \sum_{\beta \in {\cal I}}
\langle\alpha|T^{i-1}|\eta_i\rangle\sigma_i\langle\eta_i|T^{L-i}|\beta\rangle.
= \frac{1}{N^L} \sum_{\alpha \in {\cal A}} \sum_{\beta \in {\cal I}}
\langle\alpha|T^{i-1}ET^{L-i}|\beta\rangle.
  \label{eqn:1pf}
\end{equation}
where we have introduced the "evaluation" matrix
$E = \sum_{\eta_i}|\eta_i\rangle\sigma_i\langle\eta_i|$ to signify
the fact that the sum over $\eta_i$ can be expressed as a
matrix independent of the position $i$.
Similarly, the two point function is
\begin{eqnarray}
\langle\sigma_i\sigma_j\rangle&=&\frac{1}{N^L}
\sum_{\eta_i,\eta_j} \sum_{\alpha \in {\cal A}} \sum_{\beta \in {\cal I}}
  \langle\alpha|T^{i-1}|\eta_i\rangle\sigma_i
\langle\eta_i|T^{j-i}|\eta_j\rangle\sigma_j\langle\eta_j|T^{L-j}|\beta\rangle
\nonumber \\
&=& \frac{1}{N^L} \sum_{\alpha \in {\cal A}} \sum_{\beta \in {\cal B}} \left<
T^{i-1}ET^{j-i}ET^{L-j}|\beta\right>.
  \label{eqn:2pf}
\end{eqnarray}
If the power of $T$ can be evaluated,
  then it is easy to get these functions. The power of $T$ can be
  evaluated by the diagonalization of $T$. Alternatively, the direct
  multiplication of $T$ will show iterative relations which maybe easier
  to solve and will be used in the following section.

\section{$N=2$ sandpile model}
For the $N=2$ model, $T^n$ has the general form
\begin{equation}
  T^n=\left[
  \begin{array}{ccc}
    a_n & a_n & b_n \\
    a_n & a_n & b_n \\
    b_n & b_n & 2a_n+b_n
  \end{array}\right]
  \label{gfTN}
\end{equation}
The iterative relation can be obtained by $T^{n+1}=TT^{n}$, it is
$a_{n+1}=b_n$ and $b_{n+1}=2a_n+b_n$.
Therefore both of them satisfy the iterative relation
\begin{equation}
u_{n+1} = u_{n} + 2 u_{n-1}.
\label{n2}
\end{equation}
The initial condition of these
equations can be calculated from $T$,$T^2$,$T^3$ by direct
multiplication. They are $a_1=0 \;,\; a_2=1 \;,\; a_3=1$ and $b_1=1
\;,\; b_2=1 \;,\; b_3=3$. The general form of $a_n$ and $b_n$ are
$a_n=b_{n-1}=\frac{1}{3}\left(2^{n-1}-\left(-1\right)^{n-1}\right)$. Substitute
$a_n$ and $b_n$ into (\ref{eqn:1pf}),(\ref{eqn:2pf}), the results are
(for $i < j$)
\begin{eqnarray}
\langle\sigma_i\rangle&=&3/2+\left(-1\right)^i2^{-i-1}
\label{eqn:r1pf} \\
\langle\sigma_i\sigma_j\rangle&=&
9/4+\left (-1\right )^{-i+j}2^{i-j-1}+{3\,\left (-1\right )^{-i
}2^{-i-2}}+{\left (-1\right )^{j}2^{-j-1}}
\end{eqnarray}
Then two point correlation function is
(for $i < j$)
\begin{equation}
\langle\sigma_i\sigma_j\rangle_c =
\langle\sigma_i\sigma_j\rangle-
\langle\sigma_i\rangle\langle\sigma_j\rangle=
  {\left (-1\right )^{-i+j}2^{i-j-1}}+
  {\left (-1\right )^{j+1}2^{-j-2}}+
  {\left (-1\right )^{i+j+1}2^{-i-j-2}}
  \label{2pcf}
\end{equation}
These results agree with \cite{Lee1} however the derivation  here is much
simpler and more general.
The three point function can be evaluated in a similar way. We got the
three point function
(for $i < j < k$)
\begin{eqnarray}
\left<\sigma_i\sigma_j\sigma_k\right>&=&
{3\,\left (-1\right )^{-j+k}2^{j-k-2}}+
\left (-1\right )^{-k}2^{-k-1}+
{3\,\left (-1\right )^{-i+j}2^{i-j-2}}+
{9\,\left (-1\right )^{-i}2^{-i-3}}+ \nonumber \\
& &{3\,\left (-1\right )^{-j}2^{-j-2}}+
\left (-1\right )^{-i+k}2^{i-k-1}+
{\left (-1\right )^{i-j+k}2^{-i+j-k-2}}+27/8.
\end{eqnarray}
The three  point correlation function is
(for $i < j < k$)
\begin{eqnarray}
\left<\sigma_i\sigma_j\sigma_k\right>_c&=&
\left<\sigma_i\sigma_j\sigma_k\right>-
\left<\sigma_i\sigma_j\right>_c\left<\sigma_k\right> -
\left<\sigma_i\sigma_k\right>_c\left<\sigma_j\right> -
\left<\sigma_j\sigma_k\right>_c\left<\sigma_i\right> -
\left<\sigma_i\right>\left<\sigma_j\right>\left<\sigma_k\right>
\nonumber \\
&=&
{\left (-1\right )^{k} 2^{-k-3}}
+{\left (-1\right )^{k+j} 2^{-k-j-2}}
+{\left (-1\right )^{-i+k+1} 2^{i-k-2}} \nonumber \\
& & +{\left (-1\right )^{i+k} 2^{-i-k-3} }
+{\left (-1\right )^{i+k+j}2^{-i-k-j-2}}
+\left (-1\right )^{-i+k+j+1}2^{i-k-j-1}
\end{eqnarray}

Noting that  $h_i=\sum_{k=i+1}^{L} \sigma_i$,
we also get the L dependence of the correlation function of height
exactly  as follows
\begin{equation}
  \frac{1}{L}\sum_{i=1}^{L} \left(\langle h_i^2 \rangle - \langle h_i
  \rangle^2\right) =
   \frac{5L}{12}-\frac{1}{36}+\frac{2}{5L}-
  \frac{1}{9\cdot2^{2L}} - \frac{2}{9L\cdot2^{2L}} -
  \frac{1}{9\cdot2^{4L}} - \frac{8}{45L\cdot2^{4L}} -
  \frac{L}{6\cdot2^{2L}}
  \label{hcf}
\end{equation}
For large $L$, $(1/L)\sum_{i=1}^{L}\left( \langle h_i^2 \rangle -
\langle h_i \rangle^2 \right) \propto L$, which agrees with what is expected
qualitatively from the random walk argument in \cite{RWA}.

\section{$N=3$ sandpile}
For the case of $N=3$, the situation is a little more complicated. The size of
the matrix $T$ become larger. $T$ is a $4^2$ by $4^2$ matrix and the
calculation is somewhat more elaborate. The process
of solving the case of $N=3$ will show more structure of the transfer
matrix which will be useful when dealing with the cases of $N\geq3$.
The elements of $T$ for $N=3$ are $\langle ab|T|cd \rangle= \delta_{bc}
\left( 1-(1-\delta_{3a})(1-\delta_{3b})(1-\delta_{3d}) \right)$ by the
constraint (\ref{enum:constrain1}).

There is an obvious block structure of $T$ for $N=3$ model. In order
to represent this block structure succinctly, we shall adopt the
following basis which can be generalized to arbitrary $N$ later. Let
$\bf b_i$ be the column vector with component $({\bf
  b_i})_j=\delta_{ij}$.  Assign $\bf b_1 \cdots b_9$ to $|ab\rangle$,
$\bf b_{10} \cdots b_{12}$ to $|a3\rangle$, $\bf b_{13} \cdots b_{15}$
to $|3b\rangle$, and $\bf b_{16}$ to $|33\rangle$, where $a,b =0,1,2$.
The order of blocks is given by the
successive sequence of a binary expression if one replaces
3 by 1 and $a$,$b$ by 0.  The relative order
between different $|ab\rangle$s or $|a3\rangle$s, $|3a\rangle$s is not
important, but we take the order as the number sequence in base 3 for
convenience. Let $A=\{|ab\rangle | a,b= 0,1,2\}$, $B=\{|a3\rangle|
  a=0,1,2\}$, $C=\{|3a\rangle | a = 0,1,2\}$ ,and $D=\{|33\rangle\}$ to be
the subspace of original space spanned by whole vector in the basis.

The block structure for $N=3$ model becomes explicit in $T^n$,
$n\geq3$, i.e. , the matrix of $T^n$ composes of the
``saturated' block. A matrix is called ``saturated'' if all the elements
are equal. Let $ \Sigma^{m\times n}$
be the ``saturated'' matrix of $m$ rows and $n$ columns with all elements
equal to $1$. Then $T^n$ can be expressed as
\begin{equation}
  T^n=\begin{array}{r@{}l}
  &
  \begin{array}{p{2.0cm}p{2.0cm}p{2.0cm}p{2.0cm}}
    $A_9(|ab\rangle)$ & $B_3(|a3\rangle)$ &
    $C_3(|3a\rangle)$ & $D_1(|33\rangle)$
  \end{array} \\
  \begin{array}{c}
    A_9(|ab\rangle) \\
    B_3(|a3\rangle) \\
    C_3(|3a\rangle) \\
    D_1(|33\rangle)
  \end{array} &
  \left[
  \begin{array}{p{2.0cm}p{2.0cm}p{2.0cm}p{2.0cm}}
    $b_{n-2}\Sigma^{9\times9}$ & $a_{n-1}\Sigma^{9\times3}$ &
    $b_{n-1}\Sigma^{9\times3}$ & $b_{n-1}\Sigma^{9\times1}$ \\
    $b_{n-1}\Sigma^{3\times9}$ & $a_{n}\Sigma^{3\times3}$   &
    $b_{n}\Sigma^{3\times3}$   & $b_{n}\Sigma^{3\times1}$   \\
    $a_{n-1}\Sigma^{3\times9}$ & $c_{n}\Sigma^{3\times3}$   &
    $a_{n}\Sigma^{3\times3}$   & $a_{n}\Sigma^{3\times1}$   \\
    $b_{n-1}\Sigma^{3\times9}$ & $a_{n}\Sigma^{3\times3}$   &
    $b_{n}\Sigma^{3\times3}$   & $b_{n}\Sigma^{3\times1}$
  \end{array}
  \right],
\end{array} \label{eqn:bf}
\end{equation}
where $a_n$,$b_n$,$c_n$ are called block coefficients
satisfying the same iterative relation
\begin{equation}
u_{n+1}=u_n+3u_{n-1}+9u_{n-2}
\label{eqn:n3}
\end{equation}
but with different initial conditions. The initial conditions are
\begin{eqnarray}
  && a_1=0,\;a_2=1,\;a_3=4 \nonumber \\
  && b_1=1,\;b_2=1,\;b_3=4 \nonumber \\
  && c_2=1,\;c_3=1,\;c_4=7.\nonumber
\end{eqnarray}
The block structure helps in calculation. If the trivial
$\Sigma^{m\times n}$ matrix is ignored, we can use a $4$ by $4$ matrix
instead of $16$ by $16$ to represent it. This is the basic idea of
`reduced transfer matrix' which will be discussed later.

The one point function and two point function can be worked out in
the same way as the $N=2$ case.  They are
\begin{eqnarray}
\langle\sigma_i\rangle&=&2+\frac{3^{-i}\left(\omega^i+{\overline\omega}^i\right)}{2} \\
  \langle\sigma_i\sigma_j\rangle&=&
  \frac{1}{4}\left( \left[ 3^{-i}\omega^{2i-j} +
  2\omega^{i-j} + 3\omega^{-j} +
  4\omega^{-i} \right] + \mbox{h.c.} \right)+4 ,
\end{eqnarray}
where $\omega=-1+\sqrt{-2}$. The two point correlation function is (for
$i<j$)
\begin{equation}
\langle\sigma_i\sigma_j\rangle_c =
\langle\sigma_i\sigma_j\rangle-
\langle\sigma_i\rangle\langle\sigma_j\rangle=
\frac{1}{4}\left( \left[ 3^{-i}\omega^{2i-j}+
\left(2-3^{-i}\right)\omega^{i-j}-
\omega^{-j}-\omega^{-i-j}\right]+ \mbox{h.c} \right).
\end{equation}

Since the structure of the block submatrix in the transfer matrix is
simple, the original matrix can be reduced to a simpler
`reduced transfer matrix' to represent the iteration relation.  We use
capital letter to label the block, for example, $T_{AB}$
means the submatrix which maps the subset $A$ of basis into $B$.
The transfer matrix has the following
special property which produces the block
structure in $T^n$ for $n\geq3$. The sum of the elements in a row of a
submatrix, $T_{AB}$, is independent of which row one sums over. Define
the sum of the elements in any of the row of $T_{AB}$ as $\tilde T_{AB}$.
Note that we have managed to reduce each submatrix $T_{AB}$ to a number
$\tilde T_{AB}$.
Using these reduced matrices, one can simplify
the multiplication between $T$ and $T^n$ ($n\geq 3$) by inventing a
reduced matrix $R^n$ corresponding to each $T^n$ by
ignoring the trivial $\Sigma$ matrix.
Denote the block coefficient of $T^{n}_{JK}$ as $R^n_{JK}$.  Note that
$R^n$ is a 4 by 4 matrix for $N=3$ case.
One can easily show that $R^{n+1} = \tilde T R^n$.

The $\tilde T$ for $N=3$ model is
\begin{equation}
\tilde T=\left[
\begin{array}{cccc}
 0 & 1 & 0 & 0 \\
 0 & 0 & 3 & 1 \\
 3 & 1 & 0 & 0 \\
 0 & 0 & 3 & 1
\end{array}\right]
\label{eqn:rtm3}
\end{equation}
Multiply Eqn.(\ref{eqn:rtm3}) and the reduced matrix corresponding to Eqn.
(\ref{eqn:bf})
we can get the
iterative relation between $a_n$,$b_n$ and $c_n$ easily.
The reduced transfer matrix simplifies the
original 16 by 16 matrix into to 4 by 4 matrix without losing the
information of iterative relation. By reduced transfer matrix, the
iteration between $a_n, b_n, c_n$ can be written as
\begin{eqnarray*}
  a_{n+1}&=& a_n+3c_n \\
  b_{n+1}&=&3a_n+b_n \\
  c_{n+1}&=&3a_{n-1}+a_{n}
\end{eqnarray*}
It can be shown that $a_n,b_n,c_n$ satisfying Eqn.(\ref{eqn:n3}).
To solve this equation, we have to
solve the polynomial equation $x^3=x^2+3x+9$.

For arbitrary $N$, the reduced transfer matrix is a $2^{(N-1)}$ by
$2^{(N-1)}$ matrix, and they take the form
\begin{equation}
  \left[
  \begin{array}{ccccc}
    \Delta'^{N}_1 & \Delta^{N}_2 & \Delta^{N}_3 & \cdots &
    \Delta^{N}_{2^{N-2}}  \\
    \Delta^{N}_1 & \Delta^{N}_2 & \Delta^{N}_3 & \cdots &
    \Delta^{N}_{2^{N-2}}  \\
  \end{array}
  \right]
\end{equation}
where the $\Delta^{N}_i$ and $\Delta'^{N}_i$ are submatrices of the
reduce transfer matrix with $2$ columns and $2^{N-2}$ rows.
$(\Delta^N_i)_{pq}= \delta_{ip}(N\delta_{q1}+\delta_{q2})$.
$(\Delta'^N_i)_{pq}= \delta_{ip}(\delta_{q2})$. The eigenvalues of
reduce transfer matrix plays an
important role in solving the model. It can be proved that the
eigenvalues of reduced
matrix are roots of the equation $x^N=\sum_{i=0}^{N-1} N^{i}x^{N-1-i}$. The
correlation function can be expressed as the
combination of these eigenvalues. This work is in progress.

\section{Discussions}
We  have introduced the concept of transfer matrix into the study of
steady properties  of sandpile  models. This  is closely related to the
Hamiltonian formulation of the usual statistical mechanics. Indeed, one
may regard the formulas for the correlation function such as equation
(\ref{eqn:1pf}) and (\ref{eqn:2pf}) as a ``path integral'' expressions
in a discrete formulation. The transfer matrix plays the role of the
evolution operator.

The usefulness of the  transfer matrix formulation was illustrated by
deriving the one-, two- and three-point correlation functions for a
deterministic sandpile model with the critical slope $N=2$ as well as
the one- and two-point functions for the same model with $N=3$. In the
latter case, we found that the two-point function decreases exponentially
as the separation of the two point increases with  a correlation length
of $\left(\ln \sqrt{3} \right)^{-1}$  in the unit of lattice spacing. For
the $N=2$ case, the correlation length is $\left(\ln 2\right)^{-1}$. It
will be interesting  to  see if the correlation length become infinity
in the large $N$ limit so that the self-organize criticality in the
spatial correlation is restored in the limit.

This work is supported in part by  grants from the National Science
Council  of Taiwan-Republic of China under the contract number
NSC-83-0208-M001-069 and the contract number NSC-83-0208-M007-117T.

\end{document}